\documentclass[conference]{IEEEtran}
\IEEEoverridecommandlockouts
\usepackage{cite}
\usepackage{amsmath,amssymb,amsfonts}
\usepackage{algorithmic}
\usepackage{graphicx}
\usepackage{textcomp}
\usepackage{subfigure}
\usepackage{url}
\usepackage{hyperref}
\usepackage{bm}

\usepackage{listings}

\usepackage{xcolor}
\def\BibTeX{{\rm B\kern-.05em{\sc i\kern-.025em b}\kern-.08em
    T\kern-.1667em\lower.7ex\hbox{E}\kern-.125emX}}

\lstdefinelanguage{json}{
  basicstyle=\ttfamily\footnotesize,
  numbers=left,
  numberstyle=\tiny\color{gray},
  stepnumber=1,
  numbersep=5pt,
  showstringspaces=false,
  breaklines=true,
  frame=lines,
  backgroundcolor=\color{lightgray!20},
  captionpos=b, 
  literate=
   *{0}{{{\color{blue}0}}}{1}
    {1}{{{\color{blue}1}}}{1}
    {2}{{{\color{blue}2}}}{1}
    {3}{{{\color{blue}3}}}{1}
    {4}{{{\color{blue}4}}}{1}
    {5}{{{\color{blue}5}}}{1}
    {6}{{{\color{blue}6}}}{1}
    {7}{{{\color{blue}7}}}{1}
    {8}{{{\color{blue}8}}}{1}
    {9}{{{\color{blue}9}}}{1}
    {:}{{{\color{red}:}}}{1}
    {,}{{{\color{red},}}}{1}
    {\{}{{{\color{orange}\{}}}{1}
    {\}}{{{\color{orange}\}}}}{1}
    {[}{{{\color{orange}[}}}{1}
    {]}{{{\color{orange}]}}}{1},
}

\lstdefinelanguage{yaml}{
  basicstyle=\ttfamily\footnotesize,
  numbers=left,
  numberstyle=\tiny\color{gray},
  stepnumber=1,
  numbersep=5pt,
  showstringspaces=false,
  breaklines=true,
  frame=lines,
  backgroundcolor=\color{lightgray!20},
  captionpos=b, 
  literate=
    {:}{{{\color{red}:}}}{1}
    {,}{{{\color{red},}}}{1}
    {\{}{{{\color{orange}\{}}}{1}
    {\}}{{{\color{orange}\}}}}{1}
    {[}{{{\color{orange}[}}}{1}
    {]}{{{\color{orange}]}}}{1},
}

\begin{document}

\title{MPADA: Open source framework for multimodal time series antenna array measurements\\
}

\author{\IEEEauthorblockN{Yuyi Chang, Yingzhe Zhang, Asimina Kiourti, and Emre Ertin}
\IEEEauthorblockA{\textit{Department of Electrical and Computer Engineering} \\
\textit{The Ohio State University}\\
Columbus, OH 43210, USA \\
Email: chang.1560@osu.edu}
}

\maketitle

\begin{abstract}
This paper presents an open-source framework for collecting time series S-parameter measurements across multiple antenna elements, dubbed MPADA: Multi-Port Antenna Data Acquisition. 
The core of MPADA relies on the standard SCPI protocol to be compatible with a wide range of hardware platforms. 
Time series measurements are enabled through the use of a high-precision real-time clock (RTC), allowing MPADA to periodically trigger the VNA and simultaneously acquire other sensor data for synchronized cross-modal data fusion. 
A web-based user interface has been developed to offer flexibility in instrumentation, visualization, and analysis.
The interface is accessible from a broad range of devices, including mobile ones. 
Experiments are performed to validate the reliability and accuracy of the data collected using the proposed framework. 
First, we show the framework's capacity to collect highly repeatable measurements from a complex measurement protocol using a microwave tomography imaging system.
The data collected from a test phantom attain high fidelity where a position-varying clutter is visible through coherent subtraction. 
Second, we demonstrate timestamp accuracy for collecting time series motion data jointly from an RF kinematic sensor and an angle sensor. 
We achieved an average of 11.8 ms MSE timestamp accuracy at a mixed sampling rate of 10 to 20 Hz over a total of 16-minute test data.
We make the framework openly available to benefit the antenna measurement community, providing researchers and engineers with a versatile tool for research and instrumentation. 
Additionally, we offer a potential education tool to engage engineering students in the subject, fostering hands-on learning through remote experimentation.
\end{abstract}


\section{Introduction}

Vector network analyzers (VNAs) have long been the cornerstone of antenna measurements, offering unparalleled precision and reliability. 
However, as emerging technologies such as multiple-input multiple-output (MIMO) systems and RF-based biosensing evolve, the demand for multi-channel capabilities and temporal data acquisition grows exponentially. 
Existing hardware solutions often fail to meet these evolving needs, highlighting the necessity for developing frameworks that can seamlessly integrate multichannel setups with flexible time scheduling into existing antenna measurement protocols. 
This paper presents an open-source framework for collecting time series S-parameter measurements across multiple antenna elements, dubbed MPADA: Multi-Port Antenna Data Acquisition.
We make the framework available at \url{https://github.com/yuyichang/mpada}. 

Being first invented in the 1950s \cite{sayed2013vector}, VNAs have been widely used to measure reflection and transmission coefficients on a diverse range of systems such as microwave circuits, mobile base stations, and internet-of-things (IoT) devices. 
VNAs are typically used to measure antenna return signals in lieu of RF transceivers in terms of precision, dynamic range, and detailed characterization capacities. 
The ability to measure high-quality signals is especially important when the received signal consists of weak backscattered information in radar applications.
Recent studies have been focusing on inversing the unknown permittivity from extremely weak return signals mixed with antenna effects \cite{chang2023removing}.
In some cases, the weak signals are both spatially and time-varying, requiring a data collection system that attains a high level of reliability in obtaining repeatable measurements to achieve the desired signal quality. 

In recent years, there has been growing interest in multimodal signal processing. 
This developing field aims to exploit information from a variety of modalities to enhance modeling, perception, and intervention in human-machine interface (HMI) \cite{jabeen2023review}. 
RF sensing offers some unique advantages due to its privacy-preserving, noninvasive, and energy-efficient characteristics, but the ambiguous nature of the RF signal may pose challenges in processing steps which can be compensated using other sensor streams \cite{chang2024microwave}. 
As most systems are time-varying, capturing time series signals is crucial to enable a comprehensive understanding of the underlying dynamics. 
Presently, concurrent collection from VNA data and other sensor streams requires multiple data collection systems being operated independently. 
Post-processing, often done manually, is needed to align the data to the same time axis before performing further analysis. 
However, standard S-parameter measurements are not timestamped. 
The lack of timestamps hinders the RF signals from being effectively integrated into multimodal signal processing. 

MPADA is developed to address the above needs. 
The core of MPADA relies on the standard SCPI protocol, which ensures compatibility with a wide range of hardware platforms for easy integration. 
Time series measurements are enabled through the use of a high-precision real-time clock (RTC), allowing MPADA to periodically trigger the VNA and retrieve trace data. 
A web-based user interface has been developed to offer flexibility in customization by enabling control from a broad selection of devices, including mobile ones. 
The architecture also facilitates real-time monitoring to aid visualization and analysis. 
The high-precision timer also enables simultaneous data acquisition from other sensors, where the cross-modality inputs can be used for synchronized data fusion.

The remainder of the paper is organized as follows. 
Section 2 provides an overview of the software framework by discussing the design principles and individual features.
Section 3 presents data acquisition models built on top of standard S-parameter measurement procedure when time series, multi-port, and non-RF components operations are involved.
Some advanced measurement examples designed to validate the framework are provided in Section 4, focusing on bioelectromagnetics applications.
We close the paper with conclusions in Section 5.


\section{Framework overview}

\subsection{Design principles and general features}

\begin{figure}[!htbp]
    \centering
    \includegraphics[width=0.9\linewidth]{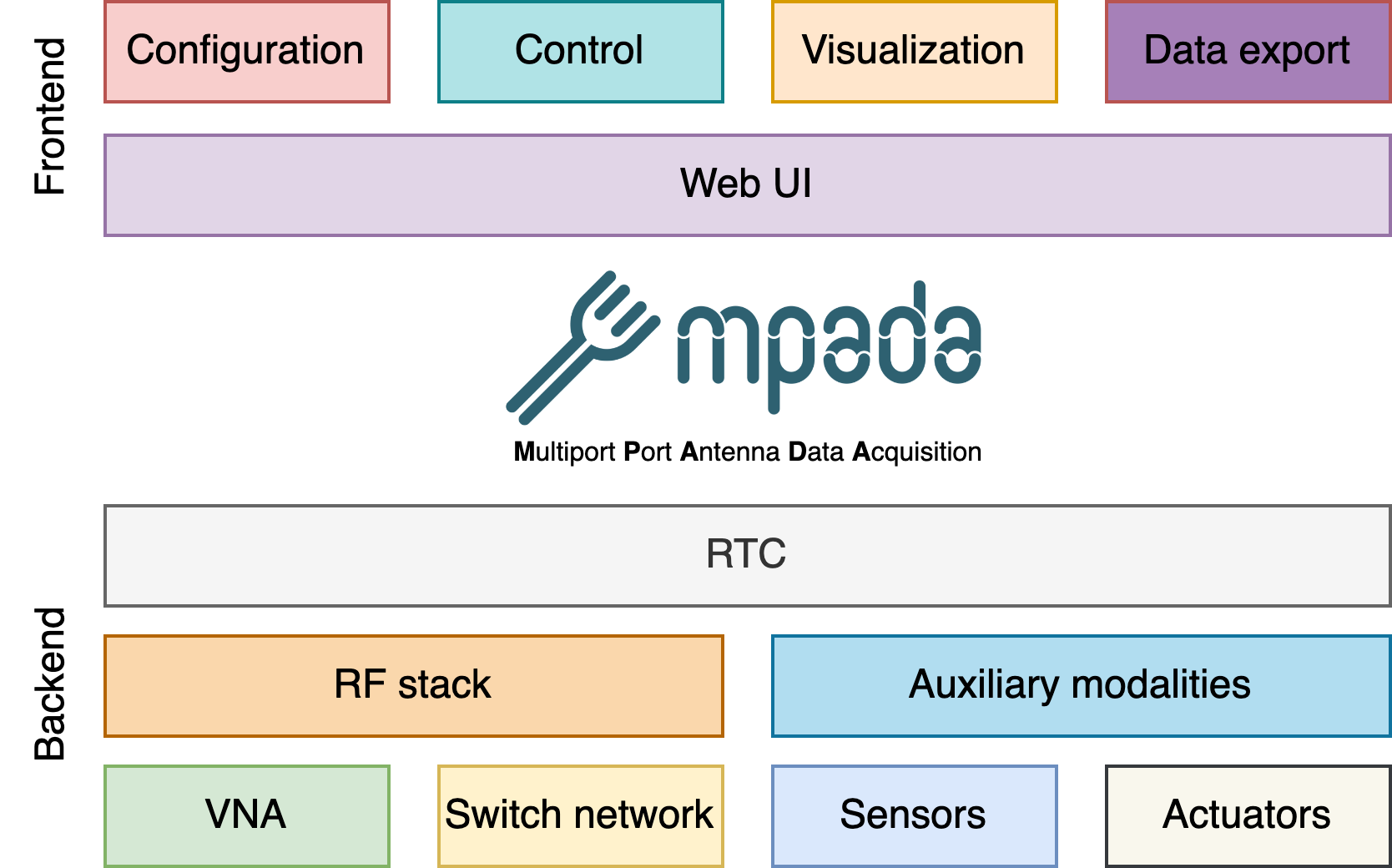}
    \caption{Architecture overview of MPADA. The front end features a web-based user interface that supports instrumentation and data input/output. The back end integrates RF and non-RF stacks enabling synchronized cross-modality data collections on the same real-time clock (RTC).}
    \label{fig:mpada}
\end{figure} 

MPADA has been developed to enable the maximum use of automation in S-parameter measurements and promote reliable and repeatable data collection.
Originally developed for collecting S-parameter data over a set sampling interval, we have designed MPADA to perform measurements that a standalone VNA is unable or difficult to achieve while considering how to integrate non-RF systems necessary for the measurement processes. 
We build the framework as cross-platform software for Linux, Microsoft Windows, and macOS. 

Python has been chosen as the primary programming language due to its significant advantages in readability, ease of maintenance, and versatility. 
Since most of the targeted users are not computer science practitioners, this makes the software more accessible and manageable for users with diverse technical backgrounds.
We employ the object-oriented programming (OOP) principle in designing MPADA. 
OOP promotes modular design where individual features of the software are managed by individual objects.
This enables easy management of software configuration and resources when enabling/disabling certain features. 
The modular nature of OOP also promotes new contributions from community developers to be easily incorporated into the main program.

Fig.~\ref{fig:mpada} provides an architecture overview of MPADA.
To date, we summarize the features that MPADA framework provides:

\begin{enumerate}
    \item Ability to automatically collect time series S-parameters across multiple RF ports;
    \item Options to enable simultaneous sensing and actuation with Non-RF hardware with set schedule;
    \item Web-based user interface that streamlines instrumentation, visualization, and data handling, accessible locally or remotely.
\end{enumerate}

\subsection{VNA communication}

The core of MPADA is its ability to instrument VNA devices via the standard commands for programmable instruments (SCPI) protocol. 
SCPI protocol is defined on top of IEEE 488.2-1987 specification that defines the syntax for command and data structure of measurement devices. 
Multiple implementations of SCPI backends exist from major device manufacturers, yet most of them are limited to certain operating systems. 
To ensure cross-platform compatibility, PyVISA-PY framework \cite{pyvisa} is used in this work as an open-source, cross-platform alternative to the proprietary libraries. 
The library is fully implemented in Python and can be installed like any standard Python package, without requiring the separate setup often needed with proprietary releases.

\subsection{User interface}

The user interface design focuses on enhancing user experience by streamlining the configuration process.
Although this is not a necessary requirement for advanced VNA users or RF practitioners, the approach aims to provide an accessible option for inexperienced VNA users, especially in education settings. 
In the classic workflow, users must navigate through multiple menus to set different components of the VNA, such as the frequency range, number of sweep points, and calibration settings. 
This "fan-out" structure can be cumbersome and inefficient, leading to potential errors and increased setup time. 
In contrast, the proposed workflow arranges all necessary configurations in sequential order on a single interface, allowing users to simply "scroll down" to complete all settings. 
This linear approach not only simplifies the process but also reduces the cognitive load on the user, making the workflow more intuitive and user-friendly. 
By minimizing the need for menu navigation, the proposed design enhances efficiency, reduces the likelihood of missed or incorrect settings, and provides a more cohesive and pleasant user experience. 
This streamlined process ensures that even users with minimal experience can efficiently configure the VNA, leading to more consistent and reliable measurements.

We use the Gradio framework \cite{abid2019gradio} when designing the user interfaces.
As a web-based user interface, the entire UI is accessible from any mainstream web browser, including mobile ones. 
This also allows users to remote access the control panels, enabling remote experimentations. 
A scripting option is also provided that allows users to quickly execute a set of commands without manually going through individual settings. 

\subsection{Data format}

The data output format of MPADA can be configured to meet different needs. 
The standard s2p format can be used to be in line with the existing VNA data format.
Additional formats like common-separated values (CSV) and byte streams (e.g., Numpy pickle file) are used to support additional data storage needs, especially when involved with non-RF data streams. 

\subsection{Peripheral interfaces}

MPADA offers the ability to interface with peripheral devices for sensing and actuation.
In many antenna measurement scenarios, sensors and actuators are needed to support data acquisition tasks yet the instrumentation is outside the scope of VNA devices.
Historically, it is common to use a separate program to perform such tasks. 
An example can be a PIC microcontroller that is used to control a rotating platform during an antenna pattern measurement. 
The setup often requires the human operator to switch between multiple programs during the measurement cycle, thus prolonging the measurement time and increasing the likelihood of error due to human factors. 
MPADA allows users to integrate controls of RF and non-RF systems into the same measurement protocol.
The measurement would then be executed automatically following the set schedule without manual intervention. 

Microcontrollers can be used as an intermediate interface when connecting a peripheral that cannot be directly plugged into the computer.
The microcontroller serves as a USB-to-UART interface since standard UART devices have become obsolete on modern computers.
This also allows the system to interface with other devices using other common communication protocols like I2C and SPI. 
The modular design of MPADA allows peripherals to be easily enabled and disabled in a configuration file, presented in Listing 1. 

\begin{lstlisting}[language=yaml, caption={Example configuration for peripheral interfaces}]
rp2040_u2if_interface_core:
  enable: True
  module: rp2040_u2if_interface.core
  object: RP2040
  label: "RP2040 U2IF Resource Manager"
tof_rp2040_u2if:
  enable: False
  module: tof_rp2040_u2if.vl53l0x.main
  object: VL53L0X
  label: "VL53L0X RP2040 U2IF"
\end{lstlisting}

\section{Data acquisition models}

In this section, we first review the typical antenna characterization workflow involving a standard S-parameter measurement. 
The difference between the VNA-only workflow and the workflow via MPADA will be discussed.
We then extend the idea to a multi-port time series S-parameter measurement case. 
Finally, we present the two prototype acquisition models for collecting multimodal measurements involving non-RF peripherals. 

\begin{figure}[!htbp]
    \centering
    \includegraphics[width=0.8\linewidth]{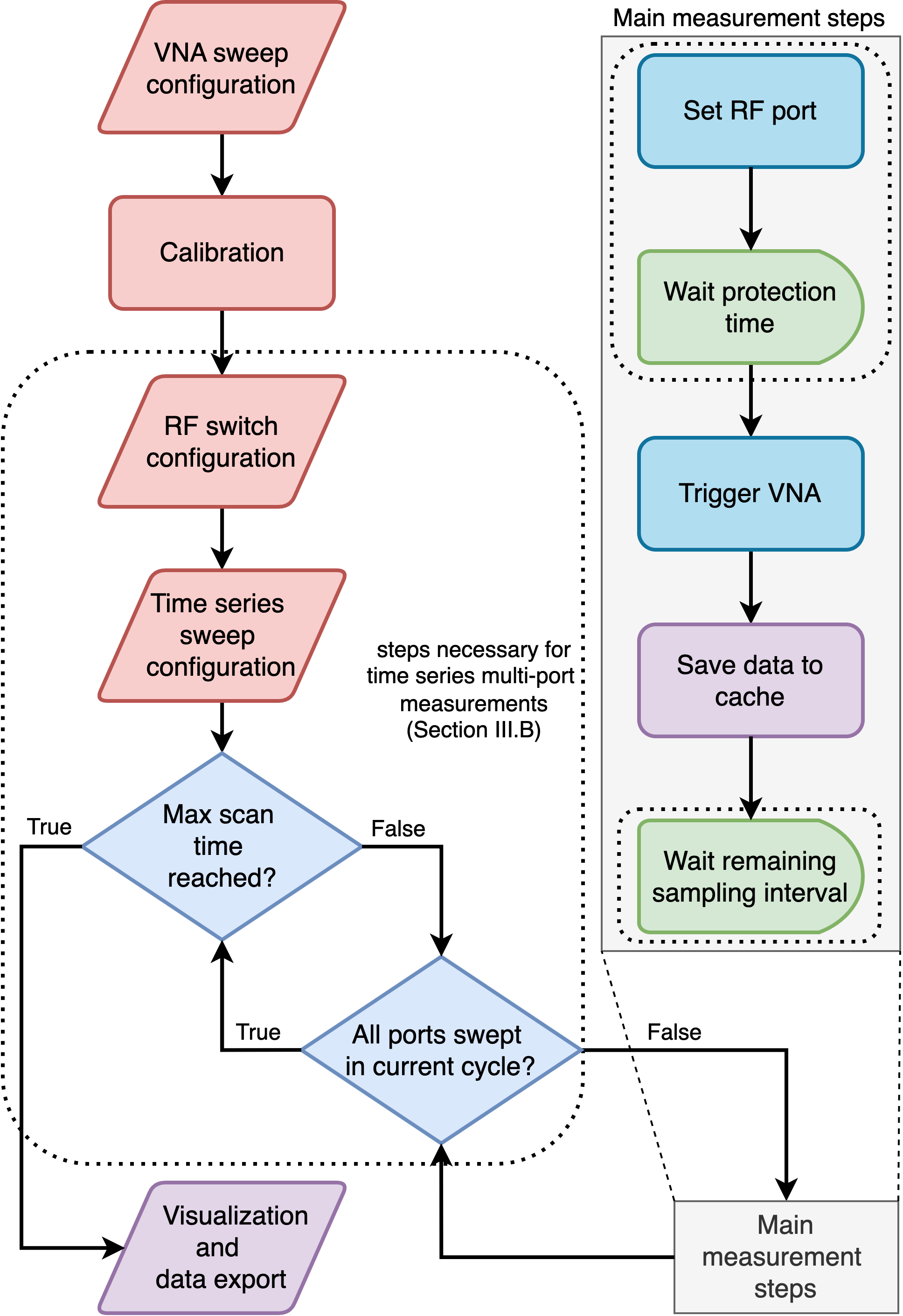}
    \caption{Measurement flowchart for collecting time series S-parameters of a multi-port system. Circled region denotes steps necessary for time series multi-port measurements.}
    \label{fig:std_workflow}
\end{figure} 

\subsection{Review of standard S-parameter measurement workflow}

Using just the VNA, the standard S-parameter measurement begins by setting up the sweep parameters like sweep frequency, the number of sweep points, and toggling traces of interests. 
A calibration step is typically performed to remove the coupling effects of non-test load components such as cables and connectors. 
Once the antenna is connected, the human operator manually triggers the VNA and saves the result to a desired location on the computer. 
If an antenna array is involved, manual configuration may be involved so that all antenna ports are measured. 

\subsection{Automation for time series antenna array measurements using MPADA}

We design a workflow that enables automatic S-parameter measurements across multiple RF ports over time, as presented in Fig.~\ref{fig:std_workflow}.
Followed by the initial setup in the standard S-parameter measurement section, the user is prompted to configure the sweep sequence of individual antenna elements. 
Listing 2 shows an example sequence where $S_{21}$ between TX1-RX1, TX2-RX2, and TX3-RX3 are measured in sequence. 
The switching between RF ports is performed by toggling the RF switch logic pin inputs. 
A pair of RF switches controlled by a microcontroller can be used, in which an example control logic is specified in Listing 3. 

\begin{lstlisting}[language=json, caption={Configuration for sweep sequence}]
{
    "1": ["TX1", "RX1"],
    "2": ["TX2", "RX2"],
    "3": ["TX3", "RX3"]
}
\end{lstlisting}

\begin{lstlisting}[language=json, caption={Configuration for RF port to GPIO pin mappings. A port is activated when predefined General Purpose Input and Output (GPIO) ports are set.}]
{
   "TX1": [],
   "TX2": ["GP0"],
   "TX3": ["GP1"],
   "RX1": [],
   "RX2": ["GP2"],
   "RX3": ["GP3"], 
   "TXcal": ["GP0", "GP1"], 
   "RXcal": ["GP2", "GP3"]
}
\end{lstlisting}

Time series sweep allows collecting time series S-parameter data by specifying a sampling interval and sampling duration.
The software triggers the VNA and timestamps the received signal using system RTC. 
The feature periodically triggers the VNA and pushes the trace data to a temporary data buffer, which is available to be downloaded at the end of the sweep. 

\subsection{Sequential and parallel modes for non-RF modalities}

\begin{figure}[htbp]
    \centering
    \subfigure[]{
        \includegraphics[width=0.46\columnwidth]{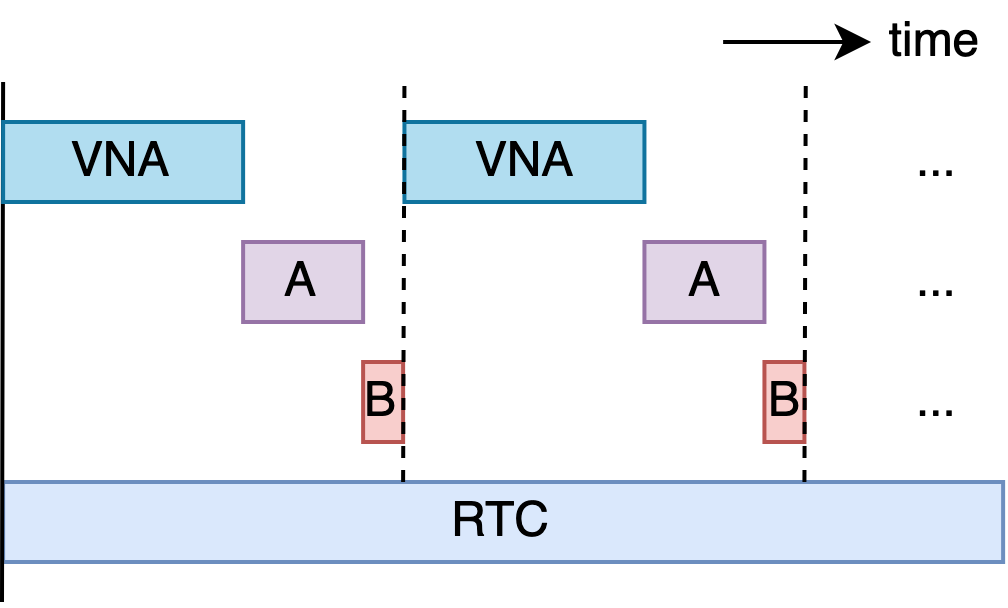}
        \label{fig:mode_s}
    }    
    \subfigure[]{
        \includegraphics[width=0.46\columnwidth]{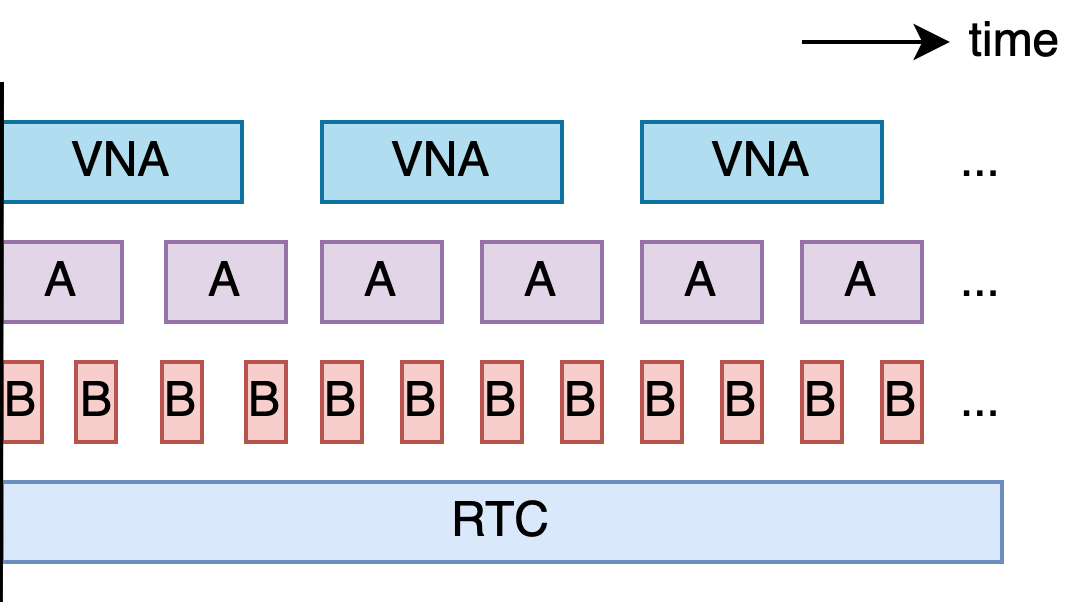}
        \label{fig:mode_p}
    }
    \caption{Two fundamental operating modes defined in MPADA for (a) sequential and (b) parallel system operations.}
    \label{fig:op_mode}
\end{figure}

We design MPADA to integrate seamlessly with non-RF sensors and actuators, enabling a synchronized way to perform multimodal data acquisition by allowing non-RF systems to operate along with the RF measurement system. 
Two fundamental operating modes are implemented as prototypes that allow users to customize the data acquisition setup based on specific experiment needs. 
In both modes, the system RTC serves as the unique timestamp reference for all modalities to ensure synchronized measurements. 
This eliminates the need to align multiple data streams in post-processing steps. 
Fig.~\ref{fig:mode_s} shows the sequential operating mode in which measurements from individual components are performed in an order orchestrated by the user. 
A specific sampling interval can be optionally set to maintain a uniform timestamp spacing. 
In contrast, parallel mode (Fig.~\ref{fig:mode_p}) allows individual modalities to operate independently under a modal-specific sampling interval. 
This is enabled through multi-processing where sub-processes are spawned to interface with specific system components.
The collected data is returned to the main process upon completion of individual sub-processes at the end of the collection time set by the user. 

\section{Advanced measurement examples}

In this section, we present some advanced use cases of MPADA with RF measurements paired with non-RF sensors and/or actuators tailored by bio-electromagnetic applications.
Using the sequential mode, we first demonstrate the robustness of the system when identifying a position-varying clutter using a microwave tomography system.  
Then, using the parallel mode, we showcase measurements from an RF kinematic sensor paired with a Hall effect sensor designed to validate the timestamp accuracy of the setup. 
Lastly, we consider using MPADA in education settings to engage students through remote experimentation.   

\subsection{Microwave tomography clutter identification by coherent subtraction}

Microwave tomography imaging is an emerging biomedical imaging modality that uses radio frequency waves to detect the change of permittivity in biological bodies. 
Based on our recent work in microwave lymphedema detection \cite{chang2024microwave}, we present measurement results from our proposed microwave tomography imaging system.
The system operates in the sequential mode where at individual angles achieved by a stepper motor, the system 1) measures transmission coefficients $S_{21}$ between individual TX-RX pairs, and 2) pulses the stepper motor to the next angle at fixed steps. 
Listing 4 presents the configuration setup where procedures specified in each row occur in sequence during a measurement cycle.

\begin{lstlisting}[language=json, caption={Configuration for microwave tomography measurements}]
{
    "sweep": "none",
    "generic_stepper.main": {"n_steps": 5}
}
\end{lstlisting}

\begin{figure}[htbp]
    \centering
    \subfigure[]{
        \includegraphics[width=0.24\columnwidth]{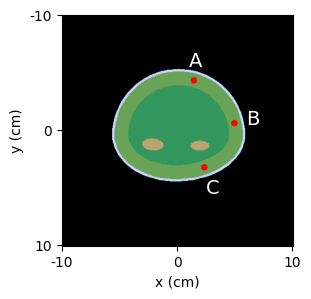}
        \label{fig:mti_pos_conf}
    }
    \subfigure[]{
        \includegraphics[width=0.68\columnwidth]{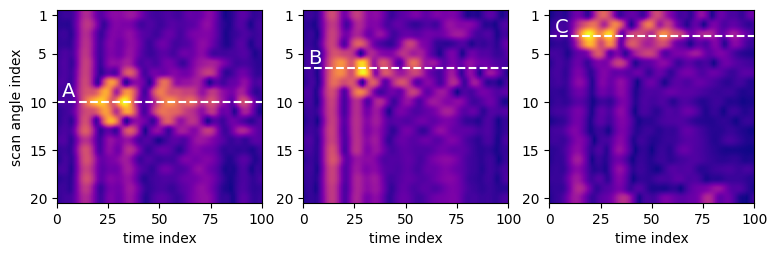}
        \label{fig:mti_pos}
    }
    \caption{Microwave tomography measurements collected using MPADA sequential mode. (a) phantom setup with red denoting lymphedema locations, and (b) $|\Delta s_{21}^p|$ after coherent subtraction at position A to C.}
    \label{fig:mti}
\end{figure}

A tissue-mimicking phantom is developed to resemble the upper extremity. 
A cylindrical-shaped lymphedema phantom is designed to be inserted within the subcutaneous region at multiple locations.
A set of tomography measurement $S_{21}^p \in \mathbb{C}^{M\times N}$ consists of $N$ frequency points averaged from all TX-RX elements collected from $M$ angles when lymphedema phantom is placed at position $p \in {A, B, C}$ shown in Fig.~\ref{fig:mti_pos_conf}. 
A scenario when no lymphedema is presented $S_{21}^{o}$ is collected and is used to coherently subtract non-lymphedema phantom contribution from collected signals.  
The time domain results are obtained by taking inverse fast Fourier transform of coherent subtraction result between cases with and without lymphedema phantom presents $\Delta s_{21}^p = \text{IFFT}(S_{21}^p - S_{21}^o)$ as shown in Fig.~\ref{fig:mti_pos}. 
Through the maximum use of automation, the measurement process requires no manual intervention for individual scan angles.
A high repeatability is achieved and is evident from changing the clutter position with respect to the actual location of the phantom of interest. 

\subsection{Timestamp verification using a RF kinematic sensor}

\begin{figure}[htbp]
    \centering
    \includegraphics[width=0.97\columnwidth]{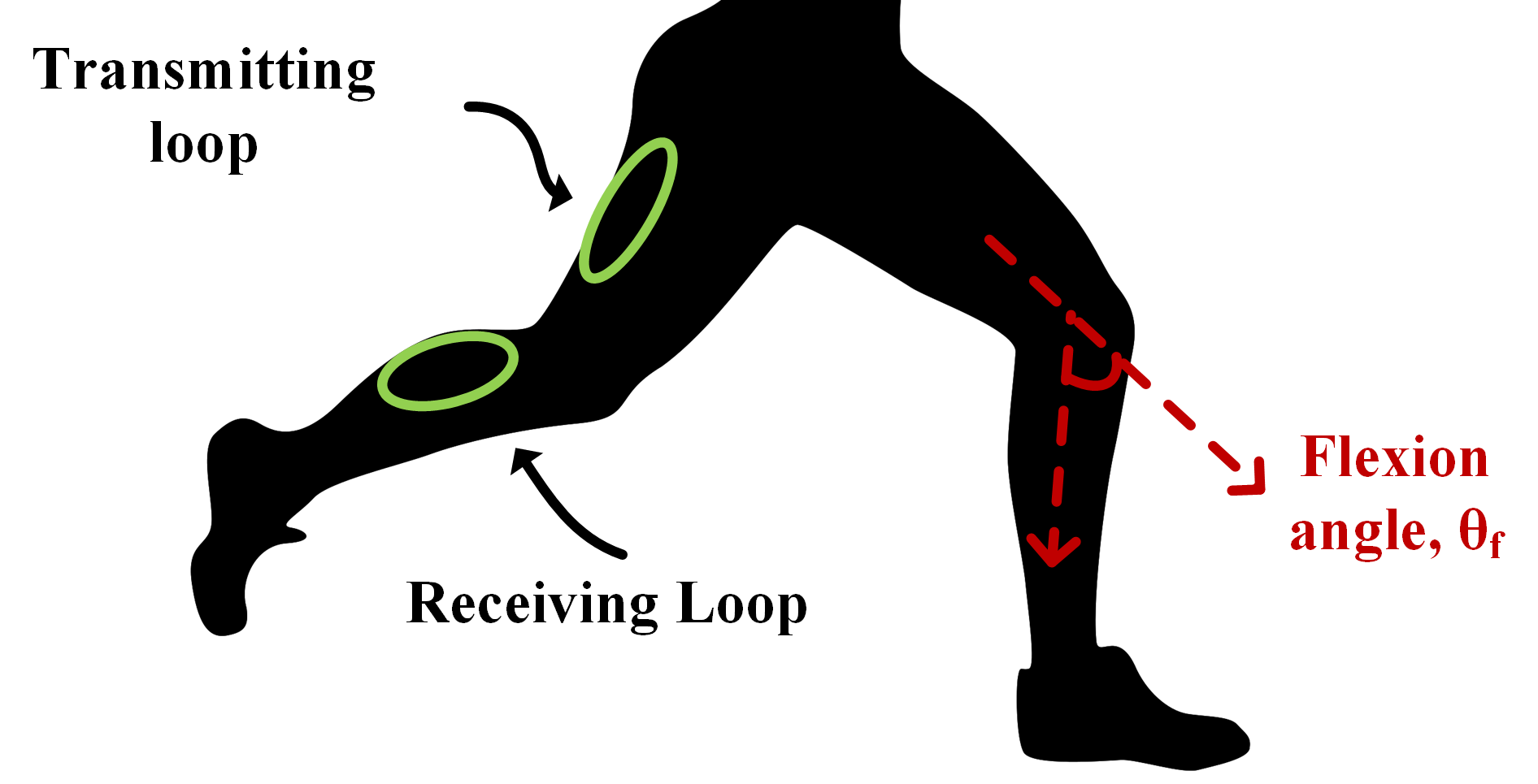}
    \caption{Schematic of the loop sensor positioned on the human leg for flexion angle measurements during body movement.}
    \label{fig:loop_sensor}
\end{figure}

Monitoring human kinematics in real-world environments has become increasingly significant in various fields, such as health care, sports, and entertainment. We recently proposed an electromagnetic-based wearable loop sensor capable of capturing human joint flexion angles in real time. The loop sensor consists of one transmitting (TX) loop and one receiving (RX) loop with a radius of 8 cm, symmetrically placed across the joint at a distance of 10 cm. An illustration of the setup is presented in Fig.~\ref{fig:loop_sensor}. 
The two loops are tuned to resonate at 34 MHz by integrating a 100 pF lumped capacitor in series. Based on Faraday's Law of induction, the power at 34 MHz from the TX loop can be coupled to the Rx loop and measured as transmission coefficient $|S_{21}|$ by a VNA. 
Joint flexion alters the relative position of the TX and RX loops, thereby changing the coupling coefficient ($\kappa$) between the loops. 
As demonstrated in our previous works \cite{8805248, s24051549}, a unique relationship exists between the joint flexion angle and $|S_{21}|$ measurements, making real-time angle estimation possible using RF-based measurements. 

\begin{figure}[htbp]
    \centering
    \includegraphics[width=0.97\columnwidth]{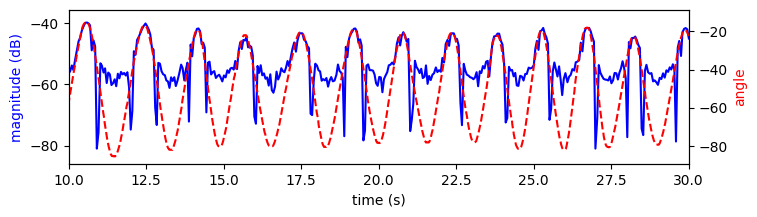}
    \caption{A 20-second snapshot of multimodal data showing $|S_{21}(t)|$ from the RF kinematic sensor (blue) and $\theta (t)$ calculated from Hall effect sensor (red) using MPADA parallel mode.}
    \label{fig:loop_data_snapshot}
\end{figure}

We consider a scenario where the RF kinematic sensor is calibrated \textit{in-situ} with a Hall effect sensor (Infineon TVL493D) that provides ground truth angle measurements. 
The Hall effect sensor is installed on one side of the phantom joint located in proximity to a radial magnet on the other side of the joint, measuring time-varying 3D magnetic flux $\bm{B}(t) = [B_x(t), B_y(t), B_z(t)]$ as a result of rotation.
As the Hall effect sensor and the magnet are attached to the different parts of the phantom, the rotation results in a change in $\bm{B}(t)$ where the angle difference can be calculated via trigonometry. 
MPADA is set to operate in the parallel mode where RF sensor $|S_{21}(t)|$ and magnetic flux $\bm{B}(t)$ are collected and timestamped simultaneously.  
A 20-second snapshot of the multimodal data collected is presented in Fig.~\ref{fig:loop_data_snapshot}.
The common time axis shared by two sensor modalities eliminates the need to align two sensor streams post-processing, enabling the potential to perform real-time signal processing. 

\begin{figure}[htbp]
    \centering
    \includegraphics[width=0.97\columnwidth]{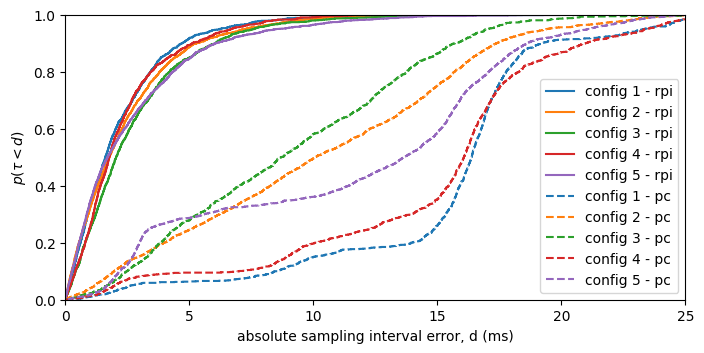}
    \caption{Empirical CDF of time stamp delay $\tau$ for individual experiment configurations running on a Raspberry Pi 4 (rpi) SBC and on a Windows 11 computer (PC).}
    \label{fig:loop_trace_tau}
\end{figure}

We verify the accuracy of the sampling interval from $|S_{21}(t)|$ timestamps collected under various configurations summarized in Table~\ref{tab:loop_conf}.
We investigate the system stability at different sampling intervals, different number of sweep points (i.e., single frequency or multiple frequency), and the way the host computer is connected to the VNA. 
The single-point sweep is collected $S_{21}(t)$ at 34 MHz, and multi-point sweeps (101) are performed between 20 to 60 MHz. 
Peer-to-peer (P2P) mode is used to directly connect the VNA and the host computer via an Ethernet cable. 
Local area network (LAN) mode is used for indirect connection tests with the use of an internet router. 
For a target sampling interval $\Bar{T}$, we calculate the actual sampling interval $T[n]$ by taking the difference between timestamps of adjacent samples where $T[n] = t[n-1] - t[n]$.
Fig.~\ref{fig:loop_trace_tau} shows empirical cumulative distribution function (CDF) curves concerning the probability of achieving the absolute sampling interval error $\tau[n] = |T[n] - \Bar{T}|$ smaller than a given threshold $d$. 
When collecting using a Single Board Computer (SBC) Raspberry Pi 4, the system achieves stable performance at $P(|\tau|<5) > 84.6\%$ and $P(|\tau|<10) > 96.5\%$. 
The results of individual configurations are presented in Table~\ref{tab:loop_conf}.
Neither the connection type nor the number of sweep points significantly influences the sampling interval delay. 
However, it is essential to ensure that the sampling interval is not set below the minimum achievable sweep time of the VNA device. 
In addition, since the framework does not operate on a Real-Time Operating System (RTOS), scheduling priority may significantly affect the sampling accuracy. 
We observe a degraded performance when the same experiments are performed on a Microsoft Windows 11 computer. 
As part of our work, we are optimizing instrumentation timing as well as porting tasks with critical time constraints to RTOS solutions. 

\begin{table}[]
\caption{Summary of experiment configurations and timestamp delay statistics obtained from Raspberry Pi 4 SBC}
\label{tab:loop_conf}
\centering
\begin{tabular}{|l|l|l|l|l|l|}
\hline
\textbf{Configuration ID}                                                         & \textbf{1} & \textbf{2} & \textbf{3} & \textbf{4} & \textbf{5} \\ \hline
\textbf{\begin{tabular}[c]{@{}l@{}}Number of \\ sweep points\end{tabular}}              & 1    & 1     & 1     & 101  & 101   \\ \hline
\textbf{Connection} & LAN & LAN & P2P & LAN & LAN \\ \hline
\textbf{\begin{tabular}[c]{@{}l@{}}Sampling interval, \\ $T$ (ms)\end{tabular}}         & 100  & 50    & 100   & 100  & 50    \\ \hline
\textbf{\begin{tabular}[c]{@{}l@{}}MSE of delay, \\ $\text{MSE}(\tau)$ (ms)\end{tabular}}      & 9.28 & 11.10 & 13.45 & 9.66 & 15.36 \\ \hline
\textbf{\begin{tabular}[c]{@{}l@{}}Variance of delay, \\ $\sigma^2(\tau)$ (ms)\end{tabular}} & 6.37 & 8.98  & 9.55  & 6.09 & 9.36  \\ \hline
\end{tabular}
\end{table}

\subsection{Curriculum development: Remote RF laboratory}
\label{sect:remote}

Laboratory plays a critical component in engineering education that provides students with hands-on experiences in reinforcing theoretical knowledge and developing practical skills.
However, there are circumstances where access to physical lab environments is limited, such as institutions with constrained resources or during events like the COVID-19 pandemic that necessitate online education. 
As a web-based application, MPADA is developed to support remote access in mind, allowing students to control the VNA and necessary peripheral devices via the internet.
This eliminates the need for physical attendance while maintaining hands-on components being exercised. 
Furthermore, using MPADA as the mid-ware serves as a security isolation that ensures the secure operation of expensive and sensitive RF devices.

\section{Conclusion}

In this paper, we present a framework to automate S-parameter measurements that is flexible to work with non-RF systems. 
Features are made to leverage the unparalleled precision offered by VNA devices to enable time series data collection across multiple RF ports. 
We demonstrate through validation experiments the system's capacity to offer highly precise and reliable measurements. 
We hope the open source nature of the framework will encourage new ideas to be added to the measurement suite and will be of future benefit to the RF research community.  

\section*{Acknowledgment}
Y. Chang and E. Ertin are partly supported by NSF Grant CBET-2037398 and NIH Grant P41EB028242. The work of Y. Zhang and A. Kiourti has been supported by NSF Grant 2042644.

\bibliographystyle{IEEEtran} 
\bibliography{bib}

\end{document}